%% file: article_4.tex
\documentclass[
10pt, 
a4paper, 
oneside, 
headinclude,footinclude, 
BCOR5mm, 
]{scrartcl}

\input{structure.tex} 

\title{\normalfont\spacedallcaps{An Adaptive Enhancement Based Hybrid CNN Model for Digital Dental X-ray Positions Classification}} 

\author{\small\spacedlowsmallcaps{{Yaqi Wang\textsuperscript{1}, Lingling Sun\textsuperscript{1}, Yifang Zhang\textsuperscript{2}, Dailin Lv\textsuperscript{1}, Zhixing Li\textsuperscript{1}, Wuteng Qi\textsuperscript{1}}}} 

\date{} 

\begin{document}


\renewcommand{\sectionmark}[1]{\markright{\spacedlowsmallcaps{#1}}} 
\lehead{\mbox{\llap{\small\thepage\kern1em\color{halfgray} \vline}\color{halfgray}\hspace{0.5em}\rightmark\hfil}} 

\pagestyle{scrheadings} 


\maketitle 
\setcounter{tocdepth}{2} 




\section*{Abstract} 

Analysis of dental radiographs is an important part of the diagnostic process in daily clinical practice. Interpretation by an expert includes teeth detection and numbering. In this project, a novel solution based on adaptive histogram equalization and convolution neural network (CNN) is proposed, which automatically performs the task for dental x-rays. In order to improve the detection accuracy, we propose three pre-processing techniques to supplement the baseline CNN based on some prior domain knowledge. Firstly, image sharpening and median filtering are used to remove impulse noise, and the edge is enhanced to some extent. Next, adaptive histogram equalization is used to overcome the problem of excessive amplification noise of HE. Finally, a multi-CNN hybrid model is proposed to classify six different locations of dental slices. The results showed that the accuracy and specificity of the test set exceeded 90\%, and the AUC reached 0.97. In addition, four dentists were invited to manually annotate the test data set (independently) and then compare it with the labels obtained by our proposed algorithm. The results show that our method can effectively identify the X-ray location of teeth.


\let\thefootnote\relax\footnotetext{\textsuperscript{1} \textit{Microelectronics CAD Center Hangzhou Dianzi University Hangzhou, China}}
\let\thefootnote\relax\footnotetext{\textsuperscript{2} \textit{West China School of Stomatology Sichuan University}}


\newpage 


\section{Introduction}
In recent years, digital apical films need electronic films with disposable envelopes. After exposure, the envelopes are removed and scanned to read. In the process, the angle of electronic dental films can not be maintained continuously. Therefore, unlike other medical images, it is necessary to adjust the angle artificially. As a result, more human errors are caused, and even medical accidents such as misdiagnosis and wrong extraction of contralateral teeth are caused. At the same time, in-depth learning technology has been widely used in the field of clinical medicine, but few studies have been applied in the field of stomatology. This study focuses on solving the problem of location and provides a technical basis for follow-up research.X-ray images are pervasively used by dentists to analyze the dental structure and to define patient’s treatment plan. However, due to the lack of adequate automated resources to aid the analysis of dental X-ray images.

In dentistry, X-rays are divided into two categories \cite{salti2002survey}: (i) Intra-oral radiographic examinations are techniques performed with the film positioned in the buccal cavity (the X-ray image is obtained inside the patient’s mouth); and (ii)extra-oral radiographic examinations are the techniques in which the patient is positioned between the radiographic film and the source of X-rays (the X-ray image is obtained outside the patient’s mouth).

Some works that use methods applied to the following types of X-ray images are analyzed: bitewing and periapical(intraoral), and panoramic (extra-oral).
Gil Silva et al. proposed the application of MASK RCNN to the data set of X-ray images outside the mouth.
\begin{figure}
	\centering
	\includegraphics[width=75mm]{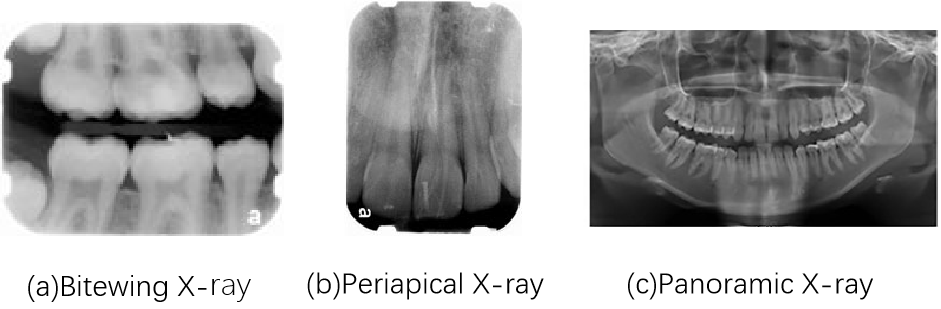}
	\caption{The various types of intraoral X-rays show different aspects of the teeth:Bite-wing X-rays,Periapical X-rays,Panoramic X-rays.}
	\label{Fig1}
\end{figure}
Abdolvahab Ehsani Rad et al.\cite{rad2015computer} applied enhancement to improve the quality of X-ray images and performed threshold processing to simplify the images. A system was developed to analyze dental X-ray images and diagnose dental caries.
Zhou and Abdel-Mottaleb \cite{zhou2005content} presented a method that consists of image enhancement, region of interest localization, and contour extraction using morphological operations and snake method. Nomir and Abdel-Mottaleb \cite{nomir2005system} developed a fully automated approach based on iterative thresholding and adaptive thresholding. 

CNN performed well in classification problems (mass or normal breast tissue classification \cite{mojahed2019convolutional}, pulmonary nodule classification \cite{van2015off}, chest X-ray classification \cite{wang2017chestx}. Therefore, the use of deep learning algorithm to analyze medical images has attracted many researchers'attention. G. Litjens and T. Kooi \cite{litjens2017survey} summarized the application of more than 300 deep learning algorithms in medical imaging. However, no literature has found that CNN is used to classify the location of dental areas. The data set of dental slices is also very scarce. In this research. CNN is used for dental X-ray regions classification.

This paper is the first time to realize the assistant recognition of teeth on apical film by means of in-depth learning. On this basis, aiming at the disorderly and arbitrary rotation of the image after scanning the bottom of the electronic apical film, the image is automatically rotated to the right angle, and the possible area of teeth in the apical film is indicated in the striking position. Assist dentists to confirm the position of teeth, reduce the possibility of misoperation of non-affected teeth, and reduce the incidence of medical accidents. At the same time, any imaging diagnosis is made up of "location + disease name". Artificial intelligence assisted confirmation of the area where the teeth are located is a necessary preparation for assisted diagnosis of oral impact.

We summarize the main contributions of this work as follows:
\begin{itemize}
	\item We propose an adaptive enhanced hybrid convolution neural network (CNN) model for classify different dental X-ray locations.  
	\item We introduce a new method to process imbalanced data in dental X-ray datasets. 
	\item We evaluate the use of CNN models in the treatment of dental X-rays of six classes, which has received little attention in previous studies. The Area Under Curve (AUC) of the test set data is 97\%, which is higher than the average evaluation results of four doctors. It shows that the method designed in this paper has good performance for unbalanced data sets. 
\end{itemize}

\section{Methods}
\subsection{Datasets}

\begin{figure}
	\centering
	\includegraphics[width=75mm]{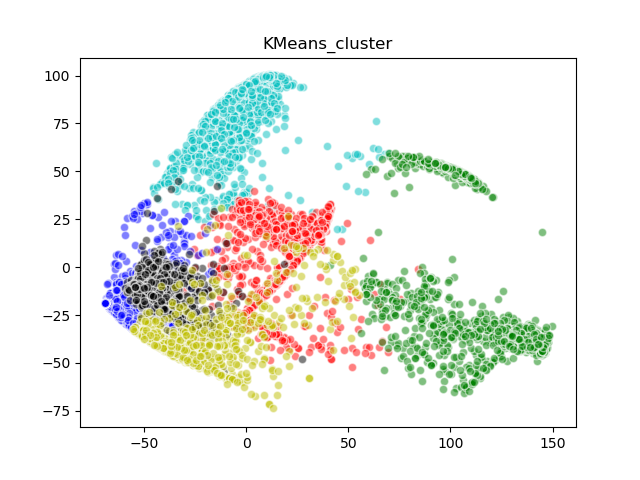}
	\caption{The Distribution of Dental X-rays Classification}
	\label{fig2}
\end{figure}

\begin{table}[]
	\caption{Six Categories of Dental X-rays}
	\begin{center}
		\begin{tabular}{@{}lll@{}}
			\hline
			Class ID   &   Class Composition  & Number \\ 
			\hline
			0        & Only teeth 11-19                                                                                    & 541   \\
			
			1        & Only teeth 21-29                                                                                & 632  \\
			2        & Only teeth 31-39                                                                              & 513   \\
			3        & Only teeth 41-49                                                              & 523   \\
			4        & Containing Class 0 and Class 1 Crossing & 242  \\
			5        & Containing Class 2 and Class 3 Crossing                                                             & 75   \\
			\hline 
		\end{tabular}
	\end{center}
	\label{tab1}
\end{table}
The purpose of this study was to achieve automatic classification of dental slice areas. Dentists classified the data sets into six categories according to the different tooth number sequences distributed on the dental slices. We used 2491 dental X-rays for training and 138 for testing.As shown in Table \ref{tab1}.
Using pHash feature extraction and kmeans image clustering, the data distribution of data set classification is shown in Figure \ref{fig2}.

In clinical diagnosis, because of the location of the film, the X-ray film of the teeth will rotate or cross, that is, the tip of the teeth will face to one side, such as 45 or 90 degrees.
In order to assist the follow-up doctor in the treatment of film reading, we not only classify the tooth regions, but also automatically correct and recognize different rotating images.

\subsection{Image Enhancement}

\begin{figure}
	\centering
	\includegraphics[width=100mm]{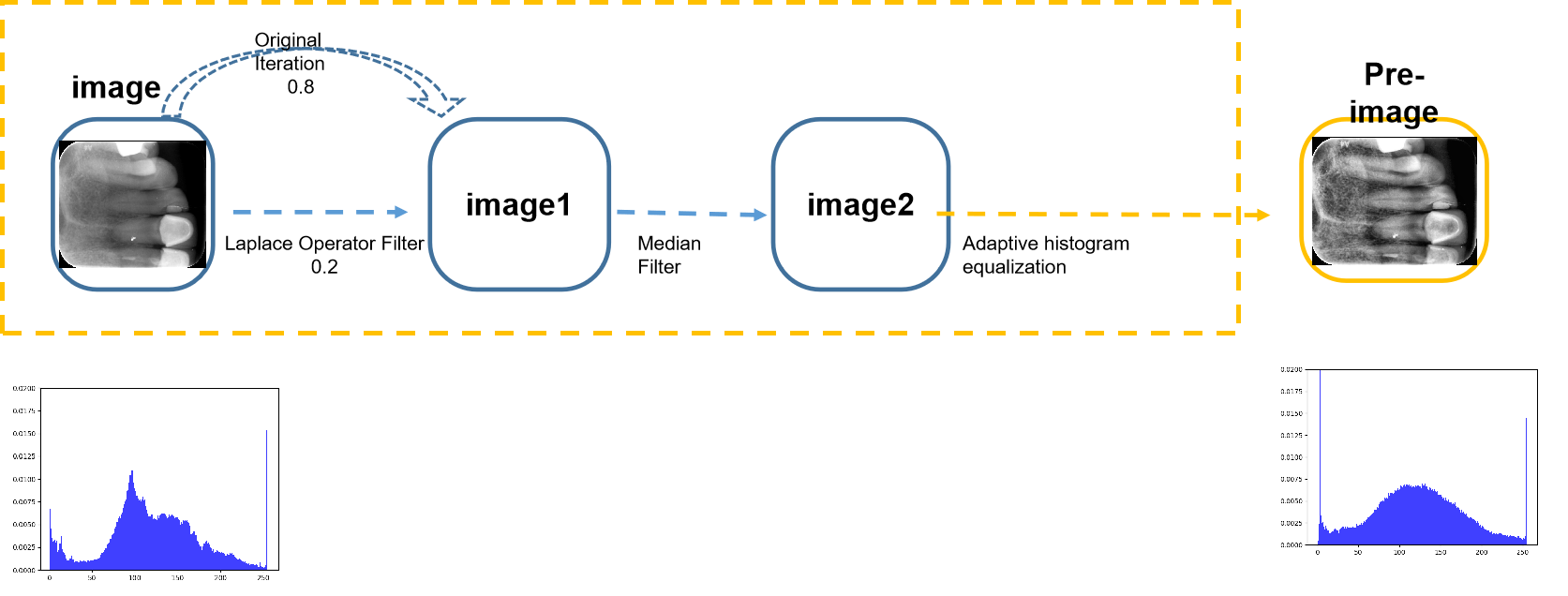}
	\caption{Data Enhancement Preprocessing Method}
	\label{fig3}
\end{figure}

Histogram equalization was used for data amplification in the category with less data (category 5 and 6).Image enhancement combines image sharpening and contrast adaptive histogram equalization. The sharpening algorithm is used to improve the details of the apex, so that the edges, contours and details of the image become clear. In this paper, Laplacian filter was used to sharpen the image. As shown in Figure \ref{fig3}. Laplace operator is a second-order differential operator commonly used in image sharpening.
Laplace transform of image function s (x, y) is defined as:
\begin{equation}
	\label{equ1}
	\bigtriangledown^{2}s=\frac{\partial^{2}s }{\partial^{2}x^{2}}+\frac{\partial^{2}s }{\partial^{2}y^{2}}
\end{equation}
Secondly, the noise in the image is filtered by median filtering method. The median filtering has a good filtering effect on impulse noise, and can also protect the edge of the image. Then, the processed image is equalized by the adaptive histogram with the limitation of contrast mentioned above. The contrast of the image is enhanced, and the artificial noise generated in the process of processing can be further eliminated. As shown in Figure \ref{fig3}.

\subsection{Network Structure}
\begin{figure}
	\centering
	\includegraphics[width=110mm]{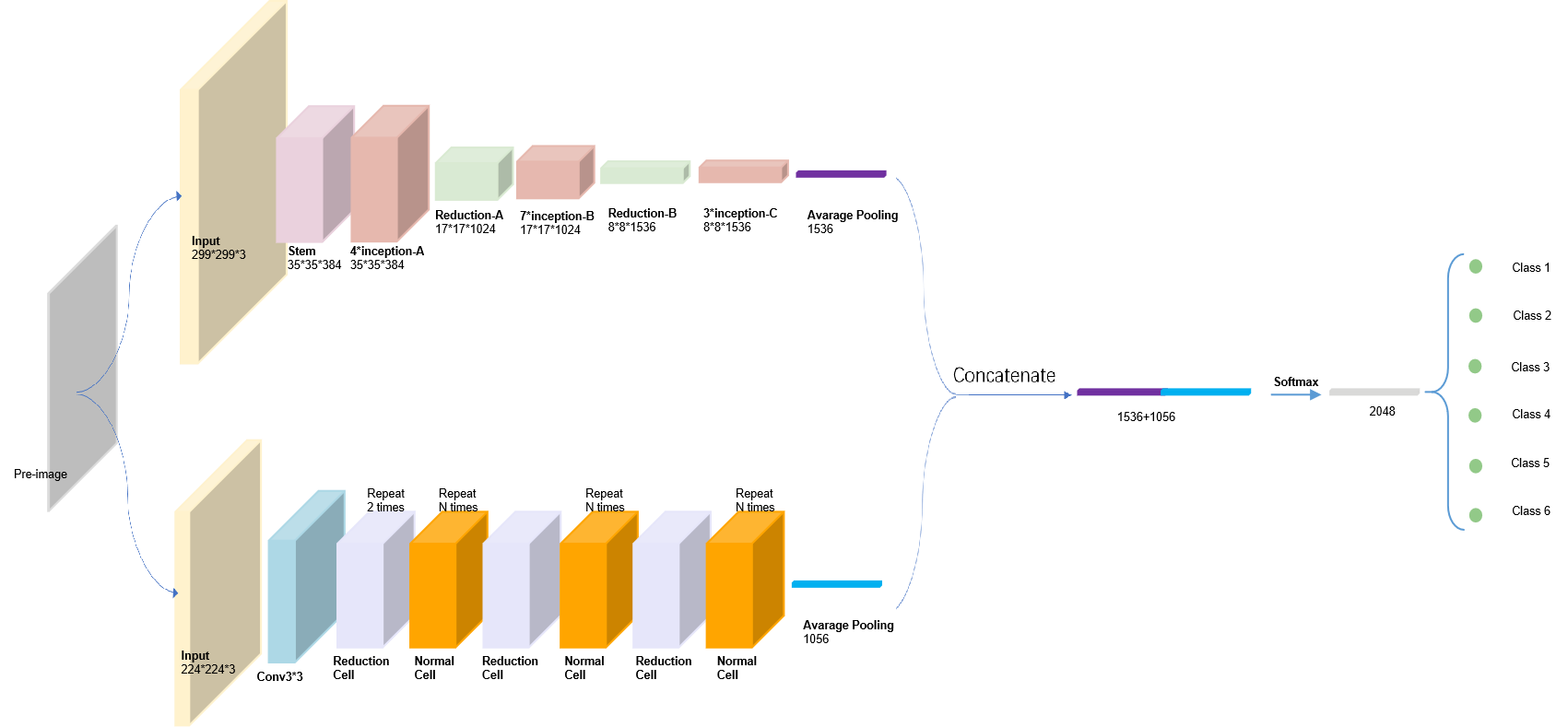}
	\caption{Mixed model of Inception V4 + NASNetMobile}
	\label{fig4}
\end{figure}

The training set was flipped (90, 180 and 270 degrees) and pre-processed by contrast enhancement, then trained with NASNetMobile and Inception V4 hybrid network model, as shown in Figure \ref{fig4}. Inception V4 + NASNetMobile: Remove the last layer of full connection (classification layer) which was trained by traditional NASNetMobile and Inception V4 networks. Each picture can generate corresponding eigenvectors after passing through two networks (NASNetMobile has 1056 features, Inception V4 has 1536 features). Then the training set, verification set and test set were extracted by this method. To save the file, A two-layer fully connected network model was built by single network training. The first layer outputs 2048 parameters and the second layer outputs six classification results. The experiment showed that adding dropout and reducing learning rate can significantly enhance the classification effect of the fusion model.

Due to the imbalance of training data categories, the weighted Ross Entropy Loss is used. The specific implementation formula is as follows:
\begin{equation}
	\label{equ2}
	loss(x,class)=weight[class](-x[class]+log(\sum _{j}exp(x[j])))
\end{equation}
Here, $j$ corresponds to the ground-truth class. 
Where $class\in[0,C-1]$ is a scalar, and is the corresponding label. And $class$ is the number of categories. $weight$ is a vector whose dimension is $class$, and representing the weight of the label. 

\section{Experiments and Evaluation}

\subsection{Evaluation Metrics}
Our test data are independent and the correct results are collected based on historical diagnostics. The test set is randomly scrambled without repeated rotation at different angles. At the same time, we invited four professional dentists to assess the regional classification of dental films. Finally, the results of the algorithm are compared with those of the doctor.

Experiments show that adding dropout and reducing learning rate can significantly enhance the classification effect of the fusion model, and the accuracy of verification set and test set can reach more than 89\%, but the accuracy of validation set and test set can not reach the highest at the same time.

\begin{figure}
	\centering
	\includegraphics[width=80mm]{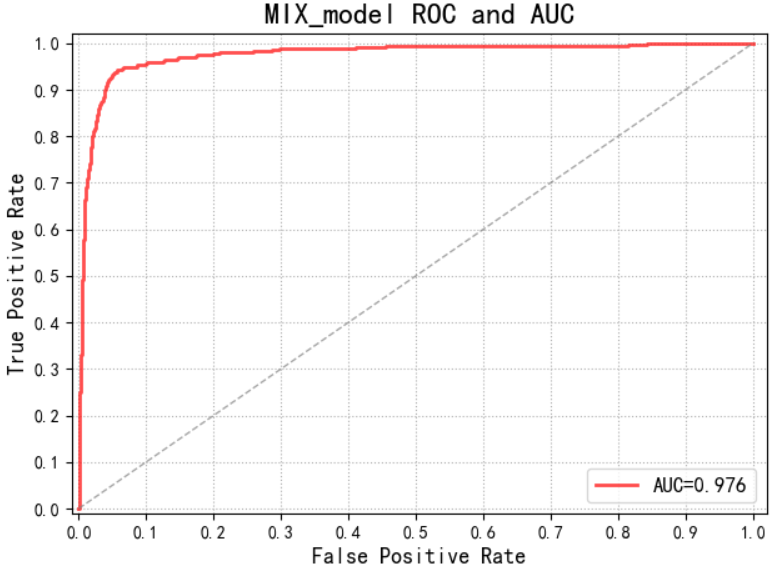}
	\caption{ROC Curve of Adaptive Enhanced Hybrid Multi-convolution Neural Network Model}
	\label{fig5}
\end{figure}

\begin{table}[]
	\caption{Six Categories of Dental X-rays}
	\begin{center}
		\begin{tabular}{@{}llll@{}}
			\hline
			Class ID   &  Precision  & Sensitivity & Specificity  \\
			\hline
			0        & 0.89 &0.77 &0.83        \\
			1        & 0.80 &0.93 &0.86       \\
			2        & 0.84 &0.90 &0.87  \\
			3        & 0.86 &0.90 &0.88  \\
			4        & 0.94 &0.89 &0.92 \\
			5        & 1.00 &0.89 &0.94  \\
			Weighted\_avg     & 0.89 & 0.88 & 0.88 \\
			\hline 
		\end{tabular}
	\end{center}
	\label{tab2}
\end{table}
The AUC value of the mixed model was 0.971, and the ROC curve was shown in Figure \ref{fig5}.

The total number of real positive categories in the sample is $TP+FN$. Similarly, the total number of real Counterexamples in the sample is $FP+TN$. $TPR$ is True Positive Rate, and $FPR$ is False Positive Rate. 

\begin{equation}
	TPR  = \frac{TP}{TP + FN};
	\label{equ3}
	FPR  = \frac{FP}{TN + FP}
\end{equation}
The ROC graph is formed by ploting $TPR$ over $FPR$, and any point in ROC space corresponds to the performance of a single classifier on a given distribution.With $FPR$ as the horizontal axis and $TPR$ as the vertical axis, the ROC curve is drawn, and the area covered downward by the curve is AUC value.

\begin{figure}[htbp]
	\centering
	\includegraphics[width=80mm]{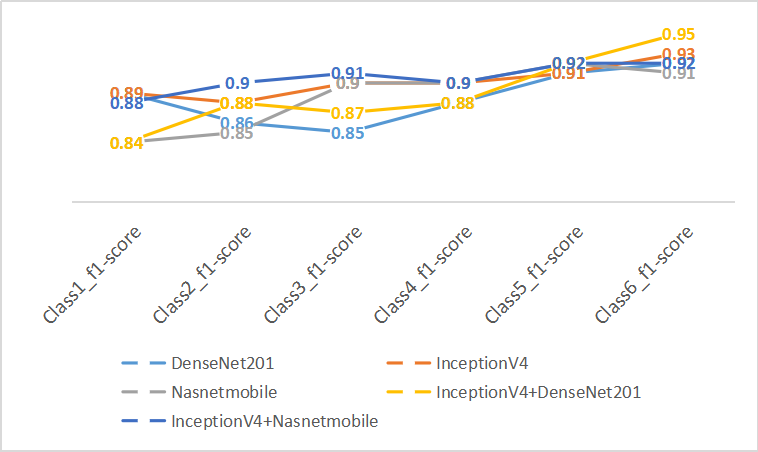}
	\caption{Comparison of recall rates of different models}
	\label{fig6}
\end{figure}


\begin{figure}[htbp]
	\centering
	\includegraphics[width=80mm]{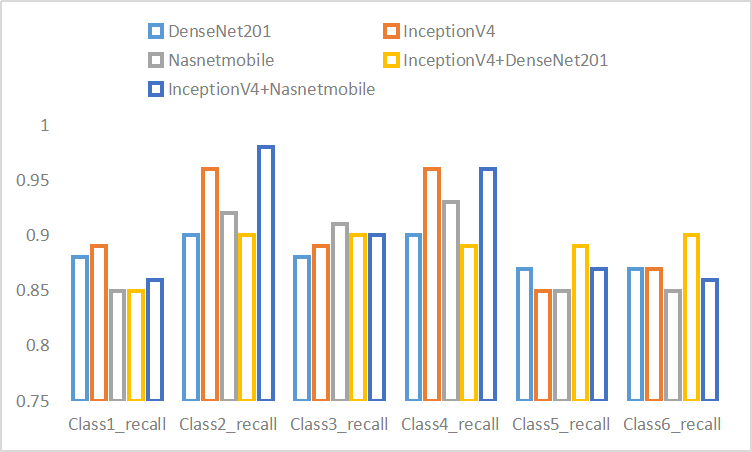}
	\caption{Comparison of F1 scores of different models}
	\label{fig7}
\end{figure}

\subsection{Evaluation and Comparison}
In addition to pretreatment of data sets, we also compared different single and mixed models, as shown in Figure \ref{fig6}f. In the process of six-classification of dental slices, we adopt many traditional classification networks to test the validity of classification network for dental slices. According to the performance of single network, we choose Inception V4, DenseNet and NASNetMobile networks based on dental film medical image data set to optimize and fuse the model. This work improves the accuracy, sensitivity and specificity of classification and prediction of dental film area.
We also made a detailed analysis and comparison of the evaluation results of the four doctors, as shown in the figure \ref{tab3}.

\begin{table}[]
	\caption{Comparison with The Average Results of Dentists}
	\begin{center}
		\begin{tabular}{llll}
			\hline
			\multicolumn{1}{l}{Models} & \multicolumn{1}{l}{Accuracy} & \multicolumn{1}{l}{Balance\_precision} & \multicolumn{1}{l}{Specificity} \\ \hline
			Doctors                      & 0.85                          & 0.87                                    & 0.85                             \\
			Muti-CNN                     & 0.87                          & 0.88                                    & 0.87 \\  \hline
			
		\end{tabular}
	\end{center}
	\label{tab3}
\end{table}

\section{Discussion}

In the process of tooth classification, there are some X-ray images of teeth with different tooth positions in the data set, so how to effectively classify them, and how to deal with the problem of tooth position intersection in a single tooth X-ray image. The preprocessing process and multi model feature fusion effectively improve the recognition rate. When blind testing data sets with several doctors at the same time, our results are better than those of doctors. This result is based on the same time dimension for test comparison. There is no doubt that if doctors are given enough time, their accuracy rate is 10\%. However, the data of the test data set is randomly scrambled after rotation, so as to simulate the non full positive emission during the dental film shooting in clinical diagnosis. During the test and comparison, we found that when doctors were asked to make decisions in a few seconds, the accuracy dropped rapidly. The data of four groups of test groups were not coincident, but the labels were consistent. As a result, our algorithm was superior to the diagnosis results given by doctors at the same time. This shows that the optimization of the algorithm is conducive to improving the diagnosis effect and efficiency under the chaotic rotation data.

\section{Conclusion}
In this paper, an automatic dental X-ray detection method based on adaptive histogram equalization and hybrid multi-convolution neural network (CNN) is proposed. The classification performance of the test set is significantly improved, and the AUC reaches 0.97. In addition, on the irregularly rotating test set, the results with four dentists showed higher efficiency and accuracy. The results show that this method can effectively identify the X-ray position of teeth and help doctors to read films. Future work will aim to use the framework of in-depth learning to segment more dental diseases and to study more methods to assist dental imaging diagnosis.



\renewcommand{\refname}{\spacedlowsmallcaps{References}} 

\bibliographystyle{unsrt}

\bibliography{ref}


\end{document}

%% file: structure.tex
\usepackage[
nochapters, 
beramono, 
eulermath,
pdfspacing, 
dottedtoc 
]{classicthesis} 

\usepackage{arsclassica} 

\usepackage[T1]{fontenc} 

\usepackage[utf8]{inputenc} 

\usepackage{graphicx} 
\graphicspath{{Figures/}} 

\usepackage{enumitem} 

\usepackage{lipsum} 

\usepackage{subfig} 

\usepackage{amsmath,amssymb,amsthm} 

\usepackage{varioref} 

\theoremstyle{definition} 

\theoremstyle{plain} 

\theoremstyle{remark} 

\hypersetup{
colorlinks=true, breaklinks=true, bookmarks=true,bookmarksnumbered,
urlcolor=webbrown, linkcolor=RoyalBlue, citecolor=webgreen, 
pdftitle={}, 
pdfauthor={\textcopyright}, 
pdfsubject={}, 
pdfkeywords={}, 
pdfcreator={pdfLaTeX}, 
pdfproducer={LaTeX with hyperref and ClassicThesis} 
}